\documentclass[twocolumn]{article}

\usepackage[top=1in, bottom=1in, left=0.75in, right=0.75in]{geometry}

\usepackage{authblk}

\usepackage{graphicx} 
\usepackage{amsmath} 
\usepackage{graphicx} 
\usepackage{hyperref} 
\usepackage{siunitx}

\usepackage[mathlines, switch]{lineno}

\title{New Directions in Focused Ion Beam Induced Deposition\\for the Nanoprinting of Functional 3D Heterostructures}

\author[1,2,3]{Frances Isabel Allen}

\affil[1]{Department of Materials Science and Engineering, University of California, Berkeley, CA 94720, USA}
\affil[2]{California Institute for Quantitative Biosciences, University of California, Berkeley, CA 94720, USA}
\affil[3]{National Center for Electron Microscopy, Molecular Foundry, Lawrence Berkeley National Laboratory, Berkeley, CA 94720, USA}

\date{October 2025}

\begin{document}


\twocolumn[
    \maketitle 
    \begin{abstract}
        The focused ion beam (FIB) microscope is well established as a high-resolution machining instrument capable of site-selectively removing material down to the nanoscale. Beyond subtractive processing, however, the FIB can also add material using a technique known as focused ion beam induced deposition (FIBID), enabling the direct-write of complex nanostructures. This work explores new directions in three-dimensional nanoprinting with FIBID, harnessing unique features of helium and neon FIBs to fabricate nanoscale  heterostructures, including multimaterial architectures and deposits with engineered internal voids. Detailed insight into the chemical and structural composition of these nanostructures is obtained using advanced electron microscopy, revealing buried interfaces and material transformations. Building on these results, the evolution of FIBID into a versatile platform for functional nanomaterials design is discussed, opening pathways toward next-generation nanoscale devices and technologies.\vspace{10mm}
    \end{abstract}
]



\section{Introduction}

Focused ion beams (FIBs) can both remove and add material with nanometric precision. Traditionally, their role as a “scalpel” has dominated, with high-resolution milling making FIBs indispensable for the cross-sectioning, trimming, and shaping of specimens for advanced microscopy and analysis, while also enabling the fabrication of novel nanostructures in fields such as photonics and plasmonics~\cite{Hoflich2023}. In contrast, the FIB's role as a “printer” via focused ion beam induced deposition (FIBID)~\cite{Utke2008} has remained comparatively underexplored, even though it offers a route to complex, functional nanoarchitectures beyond the reach of both subtractive and other additive manufacturing approaches~\cite{Jochmann2025}.

The principle of FIBID, and of the closely related technique focused electron beam induced deposition (FEBID), rests on the localized decomposition of adsorbed precursor molecules that are typically introduced via a gas injector needle. A solid deposit forms from the non-volatile reaction products of the beam-induced molecular dissociation process. While secondary electrons liberated by the impinging beam on the substrate are generally considered to be the primary drivers of the dissociation, in FIBID, especially with heavier ions, other effects also contribute, such as direct momentum transfer to the precursor molecules and energy transfer from surface atoms that are excited by the ion collision cascade~\cite{Utke2008}.

FEBID and FIBID are most commonly employed to support FIB cross-sectioning, through the deposition of uniform capping layers to protect specimens and promote even milling~\cite{Mayer2007,Uchic2007}, and for circuit edit, where conductive or insulating features can be written on demand~\cite{Orloff2003}. Yet, as alluded to above, the capabilities of these techniques extend much further, with precise beam control, sophisticated patterning strategies, a variety of precursor chemistries, and the flexibility to deposit onto non-planar substrates, making FEBID and FIBID ideally suited for the rapid prototyping of diverse functional nanostructures. Demonstrations range from Josephson junction dot arrays~\cite{Porrati2022} to complex 3D architectures such photonic chiral intertwining nanohelices~\cite{Esposito2015} and plasmonic tetragonal bipyramid structures~\cite{Winkler2017}.

Generally, FEBID has been associated with higher spatial resolution, due to the smaller probe size and more localized interaction volume for precursor dissociation~\cite{Utke2008}. However, the resolution revolution in the FIB world, brought about by the introduction of the helium ion microscope (HIM)~\cite{Ward2006}, with its sub-nanometer probe and minimal lateral scatter of the light ions within the near-surface impact zone, has meant that the resolution of FIBID can now rival that of FEBID~\cite{Alkemade2014}. For example, He-FIBID has produced nanowires with diameters of \qty{10}{nm}~\cite{Wu2014} and freestanding nanopillars just \qty{32}{nm} wide yet several microns in length~\cite{Cordoba2018}. Further applications of He-FIBID include the fabrication of superconducting nanowires and nanohelices~\cite{Cordoba2018,Cordoba2019}, atomic force microscopy tips~\cite{Nanda2015,Allen2024}, branched structures~\cite{Allen2021a,Xia2022}, and meshlike frameworks~\cite{Belianinov2020}.
Thus, even though most of the literature on nanoprinting with focused charged particle beams has emphasized FEBID~\cite{Reisecker2024}, examples of high-resolution FIBID are increasingly emerging, many of which are enabled by He-FIB technology.


As with many things in life, FIBID comes with both advantages and challenges. A notable benefit over FEBID is the higher deposition rate and more complete precursor dissociation, due to enhanced secondary electron emission and the greater transfer of both energy and momentum from the ions. However, drawbacks of FIBID are generally considered to be the co-implantation of ions and competition with sputtering. Here, He-FIBID offers unique opportunities compared to conventional Ga-FIBID. For example, by using a gaseous ion species, contamination of insulating deposits with metallic ions is avoided, and the lower sputtering power of helium enables deposition onto delicate ultra-thin substrates~\cite{Allen2024}. Even processes that are usually unwanted can be turned to advantage. An elegant example is the He-FIBID fabrication of nanopillars with hollow cores first reported in~\cite{Kohama2013}, achieved by concurrent milling by the tightly focused helium ion beam during the deposition process.

Another feature of He-FIB processing that emerges under certain conditions is the formation of helium nanobubbles in the irradiated material, which upon increasing dose can coalesce to form subsurface voids resulting in localized swelling~\cite{Livengood2009}. While usually undesired and mitigated by various strategies~\cite{Allen2021}, swelling effects can also be harnessed. Examples include kirigami-like buckling of free-standing membranes~\cite{Arora2007}, surface texturing~\cite{Zhang2015,Kim2020,Wen2022,Mo2025}, formation of subsurface nanochannels~\cite{Mo2025}, and tilting of nanostructures by local substrate swelling~\cite{Wen2022}. In He-FIBID, swelling of the deposits is typically suppressed because deposition outpaces implantation and the light ions have a long penetration depth~\cite{Kohama2013,Kohama2013b}, but if harnessed, new routes for transforming FIBID-grown architectures could emerge.


Building on the above, this article explores several new directions of FIBID that leverage some of the unique features of He-FIB processing. The first is the fabrication of nanoscale heterostructures from metallic and dielectric precursors. While there are a few reports of layered structure fabrication by alternating precursor chemistries in FEBID~\cite{Vavassori2016,Porrati2017,Keller2018}, demonstration by FIBID appears lacking. Here, it is shown that by making use of the efficient and precise patterning capability of He-FIB, and the fact that one works with a non-metallic species to keep dielectric deposits metal-free, various metal-insulator heterostructures can be realized by sequential deposition. The second exploration leverages an isotopic effect, revealed by operating the HIM ion source with neon gas~\cite{Livengood2011}. In this case, Ne-FIBID produces satellite deposits due to mass separation of the naturally abundant neon isotopes. Finally, ion-induced swelling as a means to fabricate unique nanostructures with internal voids is investigated, demonstrating the effect for both He-FIBID and Ne-FIBID. 

To delve deeper into the deposit morphologies and compositions, advanced electron microscopy is applied throughout, and potential technological applications of the novel nanostructures are discussed. In summary, this study highlights how modern FIBID, propelled by new ion sources, can evolve from a support tool into a versatile platform for functional nanoprinting of the future.

\section{Results and Discussion}

\subsection{Nanostructure arrays and multimaterial heterostructures}

\begin{figure*}[t]
    \centering
    \includegraphics[width=0.95\linewidth]{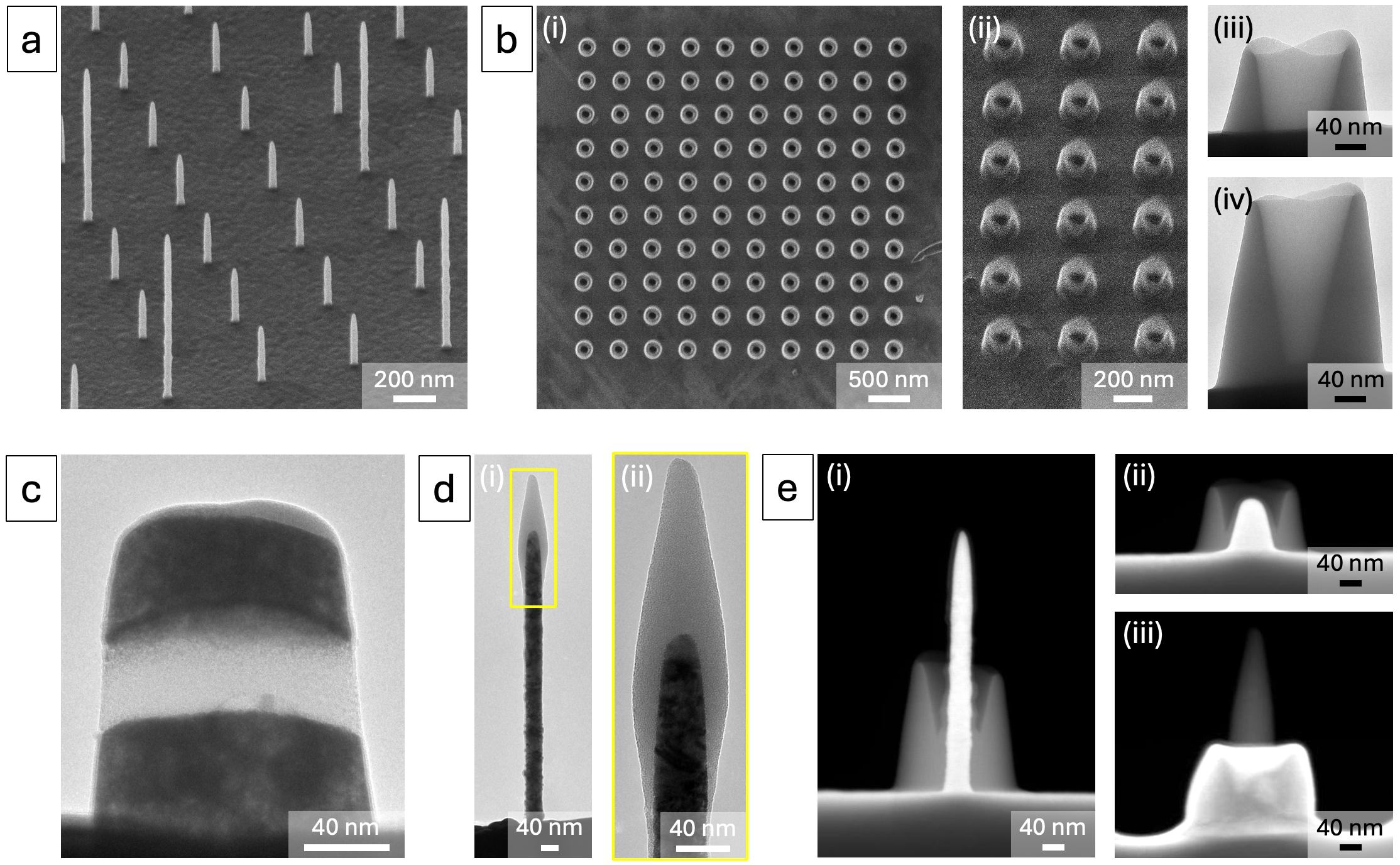}
    \caption{Examples of nanostructure arrays and individual heterostructures fabricated by He-FIBID using different precursor chemistries. (a) 54$^\circ$ tilt view HIM image of a periodic height-modulated array of metallic nanopillars. (b)(i) Plan view HIM image of an array of insulating nanorings, (ii) 40$^\circ$ tilt view HIM image of a portion of the array, (iii) and (iv) TEM side views of nanorings/nanosleeves grown to different heights. (c) TEM image of a vertically stacked metallic-insulator-metallic heterostructure, the sidewalls of which were trimmed using Ne-FIB milling after the deposition. (d)(i) TEM image of a metallic nanopillar capped with insulator, (ii) high-magnification view of the tip region. (e)(i) STEM image of a metallic nanopillar inside a shorter insulating nanosleeve, (ii) nanopillar and nanosleeve of similar heights, (iii) insulating nanopillar inside a metallic nanosleeve.}
    \label{fig:1}
\end{figure*}

Figure~\ref{fig:1} presents examples of various nanostructures fabricated by He-FIBID demonstrating the high-resolution and high-fidelity patterning capabilities of the technique, first using a single precursor chemistry per deposit and then alternating precursors sequentially to form multimaterial deposits. The precursors used were tungsten hexacarbonyl (W(CO)$_6$) and pentamethylcyclopentasiloxane (PMCPS).

The growth of single nanopillars by He-FIBID in spot mode has already been investigated in numerous works highlighting the faster growth rate compared to FEBID and superior resolution compared to Ga-FIBID~\cite{Alkemade2014}.
A further discussed advantage of the He-FIBID approach is that so-called proximity effects are negligible. That is, halo deposition from backscattered particles from the substrate is avoided, since the helium ions have a low propensity to backscatter~\cite{Alkemade2014,Schmied2015}. Through precise control of the irradiation dose, height-modulated nanopillar arrays can also be fabricated, as shown here in Fig.\ref{fig:1}(a). The widths of the nanopillars are $\sim$\qty{35}{nm}, and by implementing beam dwell times of either $\sim$\qty{3}{s} or \qty{9}{s} per point in the array, nanopillar heights of $\sim$\qty{300}{nm} and \qty{900}{nm}, respectively, were achieved. The precursor used was W(CO)$_6$, hence the pillars are metallic in character. For field-emitter applications, for example, control over the height (and pitch) of the nanopillars could be leveraged to systematically explore the dependence of tip-to-tip screening, emission distribution, and reliability on the array geometry~\cite{Harris2015,Filippov2023}. 

In Fig.\ref{fig:1}(b)(i), an example of deposition using the PMCPS precursor is shown. In this case, an array of identical nanorings was deposited, using annulus patterns with an irradiation time of $\sim$\qty{3}{s} each. The top-down view shows the inner diameters of the annuli to be $\sim$\qty{75}{nm}, with the tilt view in Fig.\ref{fig:1}(b)(ii) revealing heights of $\sim$\qty{90}{nm}. 

So far, the images shown are HIM images generated via the detection of secondary electrons due to the scanning ion beam, and thus probe the surface. In order to delve into the morphology in more detail, transmission electron microscopy (TEM) can be applied, by depositing onto TEM-compatible substrates~\cite{Allen2021a}. A result of this analysis is shown here in the 
bright-field TEM side projection views of Figs.\ref{fig:1}(b)(iii) and (iv), examining two further annulus deposits of different heights.
Nanowell structures such as these could be of interest to confine analytes or biomolecules for sensing applications~\cite{Seo2018}. 
Furthermore, given the electron transparency of the low-Z deposit material, such structures could also be used as nanosleeves for air-sensitive samples that would then be capped for inert transfer and inspection by TEM. 

\begin{figure*}[t]
    \centering
    \includegraphics[width=0.85\linewidth]{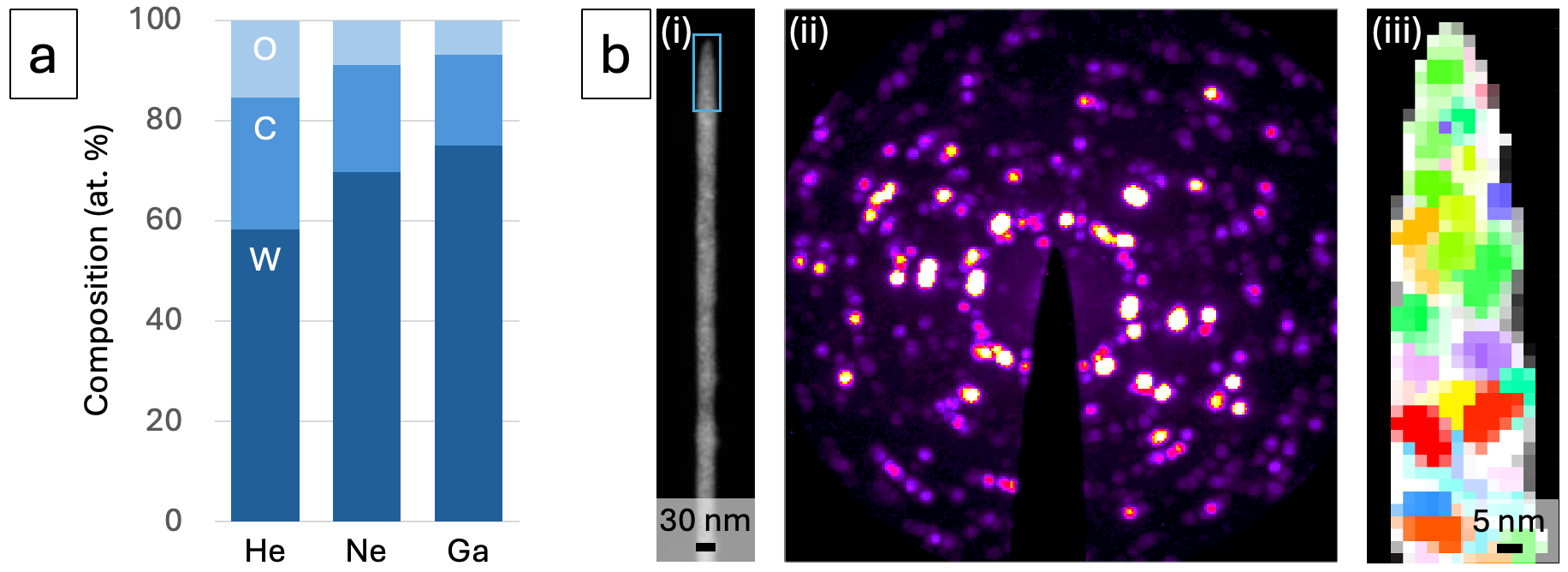}
    \caption{Elemental composition and grain size analysis of metallic deposits. (a) Relative amounts (at.\%) of W, C and O in nanostructures fabricated by He-, Ne- and Ga-FIBID from the W(CO)$_6$ precursor, determined by STEM-XEDS. (b)(i) STEM image of a nanopillar fabricated by He-FIBID from the same precursor, (ii) mean diffraction pattern computed from 4D-STEM scan of the tip apex region highlighted in the adjacent STEM image, (iii) grain map obtained following automated Bragg disk detection and non-negative matrix factorization of the 4D-STEM dataset.}
    \label{fig:2}
\end{figure*}

Now we move to the multimaterial deposits, exemplified first through Fig.~\ref{fig:1}(c). This TEM image shows a vertically-stacked trilayer structure achieved by sequential deposition of a metallic, insulating, and finally another metallic layer using the W(CO)$_6$ and PMCPS precursor chemistries. The depositions used circular patterns and the multilayer was then trimmed in-situ using a top-down Ne-FIB annulus mill. Essentially, a nanocapicitor structure has been formed. Since the ion species (He) inducing the deposition is a non-metal, the insulating layer is not contaminated with metal ions as would be the case with the more conventional Ga-FIBID. Similarly, by performing the final trim with the Ne-FIB (higher sputter yield than He-FIB due to the heavier ion), the insulating property of the central layer is preserved. Since the neon ion beam is produced by the same atomically sharp gas field ionization source (GFIS) as the helium ion beam, a small probe size is ensured, enabling the precise milling capability~\cite{Livengood2011}. 

A further example of multimaterial deposition is presented in Fig.~\ref{fig:1}(d), showing a TEM view of a metallic He-FIBID nanopillar of width \qty{40}{nm} that has been capped with insulator by top-down deposition onto the tip apex. Such a structure could be deposited onto an AFM cantilever for nanocapacitive probing. 
Along similar lines, Fig.~\ref{fig:1}(e)(i) shows a dark-field scanning TEM (STEM) image of a metallic nanopillar that was encased at the base with a nanosleeve deposited using annulus patterning with the dielectric precursor. Thus, the base of the nanopillar is electrically shielded. Such a configuration could be implemented as an insulated nanoelectrode for transmembrane intracellular probing~\cite{Fan2016}. Fig.~\ref{fig:1}(e)(ii) demonstrates tunability where both the nanosleeve and nanopillar were designed to have similar heights, while Fig.~\ref{fig:1}(e)(iii) shows the reverse arrangement, comprising a dielectric nanopillar encased by a metallic nanosleeve.

Notably, hybrid nanostructures can also be realized through post-processing techniques such as physical vapor deposition, chemical vapor deposition, and atomic layer deposition. Such approaches have been demonstrated on FEBID-fabricated scaffolds~\cite{Reisecker2024}, although in terms of nanoscale site-selectivity, in-situ multimaterial deposition, as shown here, is an attractive alternative. 

\subsection{Composition and structure analysis}

To gain more detailed insight into the composition of the deposits, elemental analysis has been performed by STEM-based X-ray energy dispersive spectrometry (XEDS). In addition, scanning nanobeam electron diffraction (also known as 4D-STEM) has been used to probe the crystallinity of the metallic deposits. 

Addressing deposition using the W(CO)$_6$ precursor, Fig.~\ref{fig:2}(a) presents results from the elemental analysis of nanopillars fabricated by He-FIBID, set in comparison with the compositions measured for Ne-FIBID and Ga-FIBID. Since these experiments were performed using a multibeam He-Ne-Ga-FIB microscope (He and Ne ions produced by the GFIS, and Ga ions produced by a standard liquid metal ion source (LMIS)), the same gas injector system was used throughout, and other key parameters such as the system base pressure, precursor flow rate, precursor cartridge temperature, and injector nozzle-to-sample distance, were kept constant. The beam energies selected were \qty{25}{keV} for He and Ne (same energy as used for all the He- and Ne-FIBID presented in this study), and \qty{30}{keV} for Ga. The beam current was set to \qty{1}{pA} for all ion species. In the compositional analysis now discussed, the contributions of implanted ions have been neglected, since the present focus is on examining differences in precursor dissociation as a function of ion mass. 

From Fig.~\ref{fig:2}(a) one can see that upon increasing the mass of the FIB species, the purity of the deposits (in terms of tungsten content) is enhanced. This trend is consistent with the findings of recent surface science studies, which attributed the enhanced metal content of heavy-ion FIBID to more efficient momentum transfer to cleave the chemical bonds within the adsorbed precursor molecules and thus liberate the organic ligands~\cite{Abdel-Rahman2024}. 
Indeed, the deposit purities from He-FIBID have been shown to be comparable to those of FEBID, and it is generally understood that Ga-FIBID, for example, produces deposits with higher metal contents than either technique~\cite{Alkemade2014}.
Nevertheless, systematic studies of these effects for multiple ion species have thus far been challenging. 
For He-FIBID applications requiring higher purity deposits, purification techniques such as those developed for FEBID could be explored, including in-situ heating or ex-situ annealing in vacuum/reactive atmosphere~\cite{Magen2021}.
Precursors designed specifically for He-FIBID with ligands that easily detach under helium ion irradiation would also be beneficial. 

The crystallinity of a metallic nanopillar deposited by He-FIBID using the W(CO)$_6$ precursor is scrutinized in Fig.~\ref{fig:2}(b), with the region selected for this analysis highlighted in the STEM image of Fig.~\ref{fig:2}(b)(i). The mean diffraction pattern obtained by scanning the selected region by 4D-STEM (scan spacing \qty{2}{nm}, 1000 diffraction patterns) is presented in Fig.~\ref{fig:2}(b)(ii) and shows many Bragg disks. This indicates the polycrystalline nature of the deposit. Following automated Bragg disk detection and grain classification by non-negative matrix factorization~\cite{Allen2021b}, the grain map presented in Fig.~\ref{fig:2}(b)(iii) was obtained. The average grain size is measured to be \qty{\sim13}{nm} with possible axial symmetry of the grain distribution revealed, suggesting radial growth from the center of the pillar in an upwards direction. In comparison, diffraction analysis of PMCPS deposits (fabricated by He-FIBID) showed them to be amorphous in nature, with elemental analysis revealing an average composition of around \qty{30}{at{.}\%} silicon, \qty{25}{at{.}\%} carbon, and \qty{40}{at{.}\%} oxygen.

\subsection{Satellite deposition}

\begin{figure}
    \centering
    \includegraphics[width=0.9\linewidth]{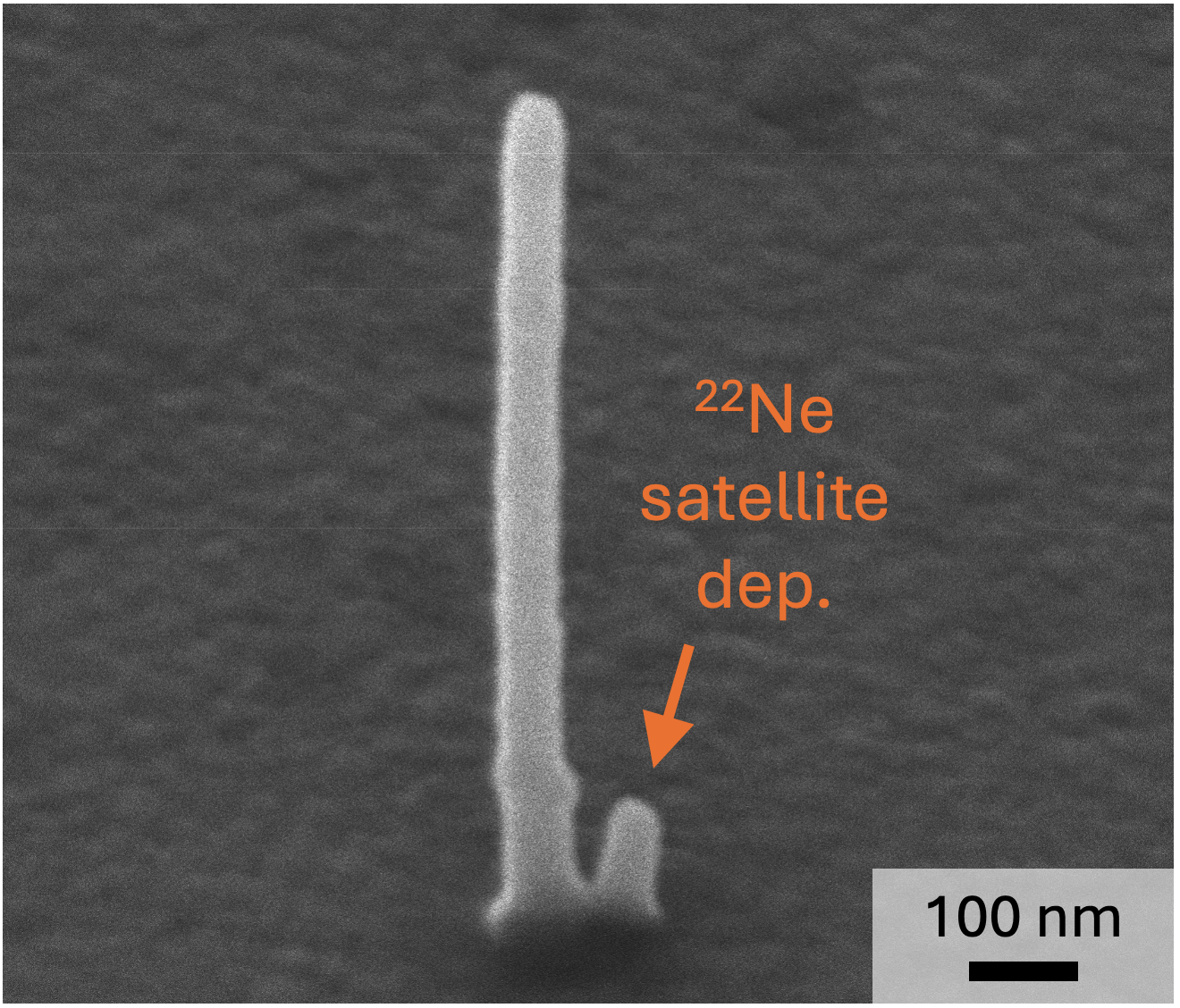}
    \caption{Metallic nanopillar fabricated by Ne-FIBID with satellite deposition due to $^{22}$Ne isotope on one side, imaged here by HIM under a tilt of 54$^\circ$.}
    \label{fig:3}
\end{figure}

In these last two subsections we take a closer look at unique aspects of He- and Ne-FIBID that might in some cases be considered undesirable, but in others could be harnessed for novel nanostructure design and application. To begin, Fig.~\ref{fig:3} presents what is in the first instance an intriguing effect, whereby a metallic nanopillar (approximately \qty{55}{nm} in diameter) was grown by Ne-FIBID and a much shorter nanopillar emerged concurrently in close proximity beside. This can be explained by an isotopic mechanism. The natural isotopic abundance of neon is \qty{\sim90}{\%} $^{20}$Ne, \qty{\sim9}{\%} $^{22}$Ne, and \qty{\sim0.3}{\%} $^{21}$Ne. Ions of different masses will follow slightly different beam paths, since for a given acceleration potential they will reach different velocities and therefore experience different focusing effects in the electrostatic lenses. Additionally, stray magnetic fields (e.g.,\ from turbomolecular or ion pumps) will deflect ions differently depending on their velocity, further contributing to mass-dependent trajectories. The satellite deposit is approximately one tenth of the height of the main nanopillar, in agreement with the abundancy ratio of $\sim$1:10 for $^{22}$Ne vs.\ $^{20}$Ne. Essentially, the volumetric ratio between the two nanopillars provides a quantitative measure of the relative abundance of the neon isotopes. 

The better known consequence of the above described isotope effect is satellite milling in Ne-FIB, and is the reason for operating the GFIS with isotopically pure neon gas in cases where this artifact cannot be tolerated. However, considering that the neon ions inevitably implant into the Ne-FIBID structures, the phenomenon observed in Fig.~\ref{fig:3} could also be exploited to achieve spatial separation and retention of isotopes at the nanoscale. 

\subsection{Nanostructures with internal voids}

\begin{figure}
    \centering
    \includegraphics[width=0.9\linewidth]{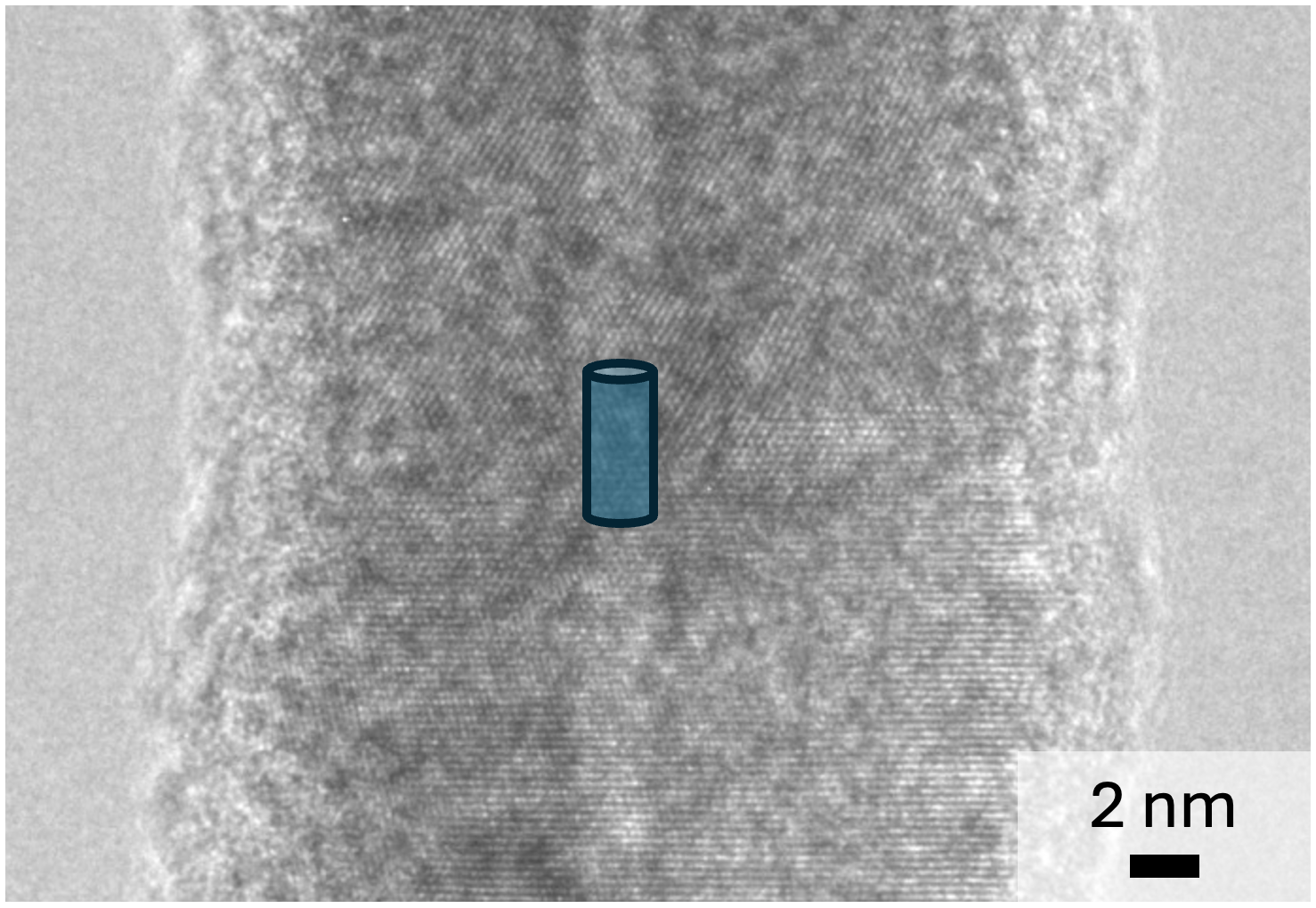}
    \caption{High-resolution TEM image of a portion of a metallic nanopillar fabricated by He-FIBID revealing a hollow core (highlighted by the blue cylinder) due to concurrent milling by the tightly focused He ion beam. This phenomenon was first reported in~\cite{Kohama2013}.}
    \label{fig:4}
\end{figure}

\begin{figure*}
    \centering
    \includegraphics[width=0.95\linewidth]{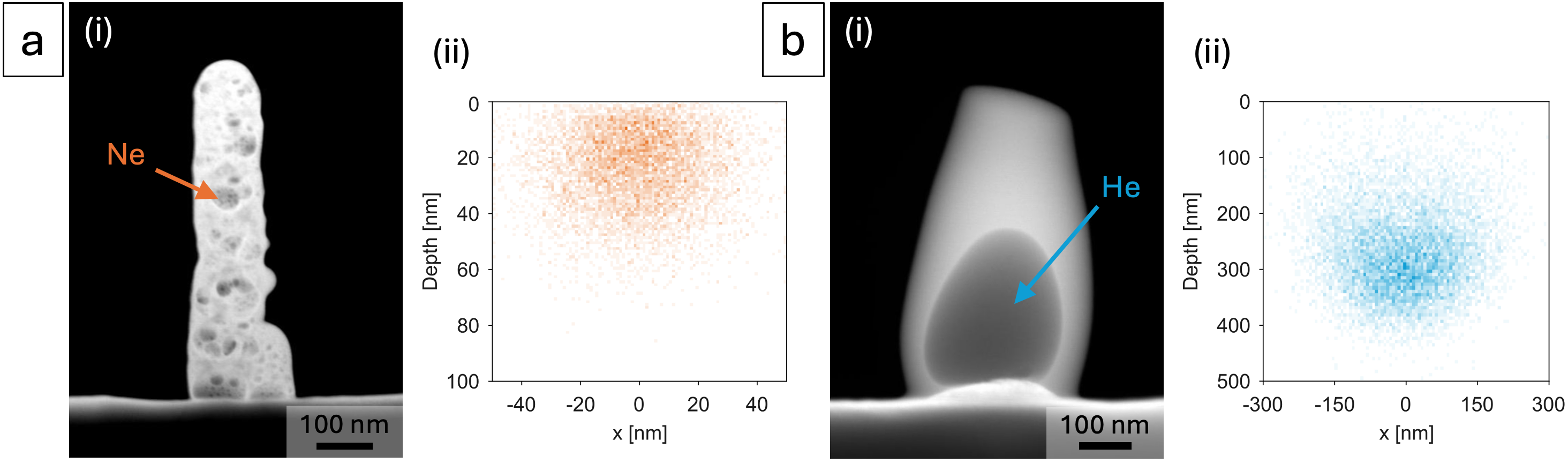}
    \caption{Nanostructures with internal voids created by ion implantation during He-FIBID and Ne-FIBID at higher doses. (a)(i) STEM image of a metallic nanopillar fabricated by Ne-FIBID with porosity introduced by Ne implantation, (ii) simulated 2D depth distribution for \qty{25}{keV} Ne ions incident on the W-C-O deposit material. (b)(i) STEM image of an insulating deposit fabricated by He-FIBID with internal void due to He implantation, (ii) simulated 2D depth distribution for \qty{25}{keV} He ions incident on the Si-C-O deposit material.}
    \label{fig:5}
\end{figure*}

Finally, we consider scenarios where voids form in deposits during the growth process. For completeness, Fig.~\ref{fig:4} highlights an effect already reported by others~\cite{Cordoba2018,Kohama2013,Cordoba2020}, which is the formation of a hollow core along the longitudinal axis of a vertically-grown He-FIBID nanopillar. The high-magnification TEM image shown here captures a $\sim$\qty{28}{nm}-wide metallic tungsten-based nanopillar with such a hollow center, with the zoomed-in inset revealing the core (in this case diameter $\sim$\qty{2}{nm}) in more detail. This core-shell phenomenon is a consequence of highly localized He-FIB milling of the nanopillar during the growth process, whereby the pillar itself is wider than the primary beam spot due to the lateral growth induced by scattered ions and their secondary electrons as they exit the sides of the structure~\cite{Chen2010}. The hollow core effect is reproduced here, since it is a fine example of how milling during FIBID can be leveraged to fabricate something novel, rather than just competing with the overall growth rate. Moreover, this phenomenon could still be exploited in future applications. For example, a hollow nanoneedle structure could be used as a nanopipette for electrochemical imaging via scanning probe microscopy, or for nanofluidic injection of individual cells or vesicles~\cite{Stanley2020}.

Larger voids can also be created, as a result of swelling induced by implanted gaseous ions during deposition at higher dose. For example, in Fig.~\ref{fig:5}(a)(i) we see a metallic nanopillar fabricated by Ne-FIBID in spot mode, with numerous internal voids revealed here by STEM. The difference between this particular deposition and that of the narrower Ne-FIBID nanopillar shown in Fig.~\ref{fig:3} was the beam current used: \qty{0.5}{pA} for the former nanopillar and \qty{1}{pA} for this wider one with voids. Increasing the beam current slowed the vertical growth rate, indicating a precursor-limited regime in which excess ion flux drives void formation due to the implantation of neon. To the right of the nanopillar one can again discern satellite deposition from $^{22}$Ne, manifesting here as a broader shoulder-like feature due to the overall wider deposition profile. 

The depth distribution of the voids can be understood from simulations of the ion stopping range in the deposit material, as shown in Fig.~\ref{fig:5}(a)(ii). Plotted are the final resting points (in coordinates of depth vs.\ lateral scatter) of \qty{25}{keV} neon ions in the W-C-0 deposit material using Stopping and Range of Ions in Matter (SRIM) code~\cite{Ziegler2010}, assuming the material composition measured previously by STEM-XEDS, and irradiation of a single point. The concentration of implanted ions from the simulation is greatest at a depth of $\sim$\qty{20}{nm}, hence voids can be expected to form relatively close to the surface as the structure grows, resulting in the fairly even final distribution observed. Porous nanostructures such as these are of interest for their large surface area and mechanical properties, with applications ranging from catalysis and sensing to metamaterials combining low mass-density with high strength~\cite{Li2024}. 

Figure~\ref{fig:5}(b)(i) shows another example of void formation, in this case for He-FIBID. This pillar structure was fabricated using the PMCPS precursor and a relatively high irradiation dose, patterning a disk-shaped region of diameter \qty{200}{nm}. The STEM image reveals a single void occupying most of the lower half of the deposit, which presumably formed as a result of the coalescence of helium nanobubbles, as has been shown for helium ion irradiation of a tungsten target elsewhere~\cite{Allen2020}. Simulation results from SRIM for the ion stopping range are shown in Fig.~\ref{fig:5}(b)(ii), computing for \qty{25}{keV} helium ions using the Si-C-O deposit composition obtained by STEM-XEDS. Here, most ions accumulate at a depth of $\sim$\qty{300}{nm}, so much deeper than in the Ne-FIBID case, driving void formation in the lower portion of the He-FIBID deposit. In fact, a degree of substrate swelling is also visible. The simulation also shows that lateral straggle of the implanted ions is greater in the He-FIBID case, which may help explain the formation of the single void. 
A nanocontainer structure with a well-defined cavity such as this could enable studies of the pressurized encapsulated gas, for example, by electron energy-loss spectroscopy~\cite{Taverna2008}, or even serve as a nanoscale reaction vessel. Post-deposition, additional voids could also be introduced at targeted locations by adjusting the beam energy and angle of incidence, enabling the formation of tailored multi-compartment nanostructures.


\section{Conclusion} 

This study demonstrates that FIBID using finely focused helium and neon ion beams enables the efficient and versatile fabrication of heterostructures in the \qty{100}{nm} size range and below. Sequential deposition of electrically conductive and insulating layers is achieved without contaminating the dielectric components by metal ions, while gaseous ion-induced swelling provides an additional route to structural control. In addition, by exploiting mass-dependent ion trajectories, the separation of isotopes into individual nanostructures is shown. Advanced electron microscopy offers detailed insight into deposit composition and interface morphology, with grain mapping by 4D-STEM serving as a valuable tool for future efforts aimed at tuning crystallinity through post-processing. 

The sequential multimaterial strategies shown here can be directly extended to FEBID, albeit with the slower growth rates inherent to the electron beam technique, whereas the formation of internal voids represents a unique capability of gaseous-ion FIBID. Looking ahead, these approaches open opportunities for applications ranging from nanoelectrodes for transmembrane biosensing to nanoscale reaction vessels for confined chemical reactions, as well as for the post-growth manipulation of nanostructures by ion implantation.

\section{Experimental Section} 

The FIBID experiments were performed using a Zeiss ORION NanoFab microscope, which combines a specialized GFIS for He and Ne ions with a conventional LMIS for Ga ions. A gas injector system (GIS) from Oxford Instruments, the OmniGIS II, was used to deliver the gaseous precursors W(CO)$_6$ and PMCPS, with flow rates adjusted to target chamber pressures of \qty{2e-6}{Torr} or \qty{8e-6}{Torr} for the metallic precursor, and \qty{2e-6}{Torr} for the dielectric. The base chamber pressure was \qty{1e-7}{Torr}. The W(CO)$_6$ cartridge was heated to \qty{50}{\degree C}, whereas the PMCPS cartridge was not actively heated. For the sequential deposition sequences, base pressure was restored before switching to the next precursor chemistry. The distance from the end of the injector needle to the sample was set to $\sim$\qty{100}{\micro\meter}.

For He-FIBID, the nominal probe size was \qty{0.5}{nm}, selecting a \qty{10}{\micro\meter} aperture and spot 4. For Ne-FIBID, the nominal probe size was \qty{2}{nm}, using a \qty{20}{\micro\meter} aperture and spot 4. He beam currents ranged from \SIrange{1}{3}{pA} and Ne beam currents ranged from \SIrange{0.5}{2}{pA}. The He and Ne beam energies were set to \qty{25}{keV}. For the comparative Ga-FIBID results presented in Fig.~\ref{fig:2}(a), a Ga beam current and energy of \qty{1}{pA} and \qty{30}{keV}, respectively, were selected.

Nanopatterning was performed using NanoPatterning and Visualization Engine (NPVE) software from Fibics, Inc. Nanopillars were patterned in continuous-dwell spot mode using doses of \SIrange{10}{100}{\micro\coulomb\per\micro\meter\squared}. Annuli and filled circles were patterned using a beam overlap of \qty{50}{\%} and a dwell time per pixel of \qty{1}{\micro\second}. The annulus patterns (circular scans) had a diameter of \qty{150}{nm} and ring thickness of \qty{1}{nm}, and used doses of \SIrange{5}{20}{\nano\coulomb\per\micro\meter\squared}. The filled circle patterns were scanned line-by-line in serpentine or standard mode, with further parameters detailed case-by-case below.
HIM imaging (top-down and under stage tilt) was performed by secondary electron detection. The substrates used were either bare copper or \qty{30}{nm} gold on borosilicate glass, and in the case of deposits inspected by TEM/STEM, custom FIB-milled copper half-grids~\cite{Allen2021a}.

Analysis of the deposits by TEM and STEM was performed using an FEI TitanX electron microscope operated at \qty{80}{kV} and \qty{300}{kV}. For the elemental mapping, an FEI Super-X quadrature X-ray detector was used together with Bruker Esprit software. 4D-STEM datasets were acquired using custom scripts and analyzed using custom MatLab code for automated Bragg disk detection with grain classification performed using non-negative matrix factorization~\cite{Allen2021b}. The electron probe full-width-at-half-maximum was estimated at \qty{2.4}{nm} and the scan step size implemented was \qty{2}{nm} with a dwell time of \qty{0.05}{s} per pixel. The rectangular region shown in Fig.~\ref{fig:2}(b)(i) had a scan size of 20$\times$50 pixels, corresponding to 1000 diffraction patterns. 

Additional fabrication details relating to the heterostructure deposits are given below.

Fig.~\ref{fig:1}: (1) The multilayer structure in Fig.~\ref{fig:1}(c) was deposited sequentially by He-FIBID using disk (filled circle) shapes of diameter \qty{200}{nm} with a dose of \qty{3}{\nano\coulomb\per\micro\meter\squared} for the metallic layers (at \qty{1.8}{pA}) and \qty{0.5}{\nano\coulomb\per\micro\meter\squared} for the central dielectric layer (at \qty{2}{pA}). The structure was then trimmed by Ne-FIB top-down milling using a \qty{1}{pA} beam and an annulus pattern with inner and outer diameters of \qty{150}{nm} and \qty{300}{nm}, respectively, using a dose of \qty{0.5}{\nano\coulomb\per\micro\meter\squared}. 
(2) The metallic nanopillar in Fig.~\ref{fig:1}(d) was capped top-down by He-FIBID in spot mode with PMCPS using a dose of \qty{10}{\micro\coulomb\per\micro\meter\squared} (beam current \qty{2}{pA}).
(3) The nanopillar-in-nanosleeve structures in Figs.~\ref{fig:1}(e)(i) and (ii) were created by growing the nanopillars first, whereas for Fig.~\ref{fig:1}(e)(iii) the nanosleeve was grown first.

Fig.~\ref{fig:3}:
The Ne-FIBID nanopillar with well-defined satellite was deposited in spot mode from the W(CO)$_6$ precursor using a beam current of \qty{0.5}{pA} and a dose of \qty{100}{\micro\coulomb\per\micro\meter\squared}. 

Fig.~\ref{fig:4}:
The He-FIBID hollow nanopillar was deposited in spot mode from the W(CO)$_6$ precursor using a beam current of \qty{1}{pA} and a dose of \qty{40}{\micro\coulomb\per\micro\meter\squared}. 

Fig.~\ref{fig:5}:
(1) The Ne-FIBID porous nanopillar in Fig.~\ref{fig:5}(a)(i) was deposited in spot mode from the W(CO)$_6$ precursor using a beam current of \qty{1}{pA} and a dose of \qty{240}{\micro\coulomb\per\micro\meter\squared}. 
(2) The He-FIBID nanostructure with single internal void in Fig.~\ref{fig:5}(b)(i) was deposited using a circular pattern of diameter \qty{200}{nm} from the PMCPS precursor with a beam current of \qty{2.5}{pA} and dose of \qty{5}{\nano\coulomb\per\micro\meter\squared}, i.e., a factor of 10 higher dose than that used for the deposition of the He-FIBID dielectric layer shown in Fig.~\ref{fig:1}(c).

\section*{Acknowledgements} 

The author thanks John Notte for helpful discussions and Thomas Pekin for assistance with the preliminary 4D-STEM analysis. Ion microscopy and nanofabrication was perfomed at the Biomolecular Nanotechnology Center, a core facility of the California Institute for Quantitative Biosciences at UC Berkeley. Electron microscopy was performed at the Molecular Foundry, which is supported by the Office of Science, Office of Basic Energy Sciences, of the U.S. Department of Energy under Contract No.\ DE-AC02-05CH11231.

\bibliographystyle{unsrt} 
\bibliography{bib} 

\end{document}